\documentclass[twocolumn,prd,aps,showpacs]{revtex4}
\def\be{\begin{equation}}
\def\ee{\end{equation}}
\def\ba{\begin{eqnarray}}
\def\ea{\end{eqnarray}}
\def\half{{1 \over 2}}

\def\nave{n_{\rm ave}}
\def\dn{\delta n}

\def\tpsi{\psi}

\def\dt{\Delta t}
\def\dx{\Delta x}
\input epsf
\begin{document}
\title{Classical stability of supercurrent in one dimension: a numerical study}
\author{S. Khlebnikov}
\affiliation{Department of Physics, Purdue University, West Lafayette, 
IN 47907, USA}
\begin{abstract}
We report results of a classical simulation of thermal phase slips, and the
associated relaxation of supercurrent, in a ring-shaped one-dimensional
superfluid. We find that the {\em classical} relaxation rate vanishes in 
the uniform limit. This leaves the {\em quantum}
relaxation, with momentum transfer to phonons, the only mechanism of supercurrent
decay in the uniform system. In the presence of a smooth periodic potential, 
classical decay becomes possible, and we identify a family of moving critical 
droplets that can mediate it.
\end{abstract}
\maketitle
\section{Introduction}
Decay of supercurrent in one-dimensional (1D) superfluids and superconductors is
a problem of longstanding interest. In 1967, Little \cite{Little} discussed the 
role of ``large'' thermal fluctuations in thin superconducting wires and observed
that if, as a result of such a fluctuation,
the order parameter $\Psi$ vanishes at some point, the phase of $\Psi$ can unwind, 
leading to a nonzero resistance. Such processes became known as thermally-activated
phase slips (TAPS). Using thermodynamic arguments, Little had pointed out that at low 
temperatures the rate of these processes is exponentially small.

A more detailed theory of TAPS, based on the Ginz\-burg-Landau (GL) equations, was 
constructed in 
subsequent works \cite{LAMH} and became known as the LAMH theory. At first sight,
it may appear that the LAMH theory is essentially an application of the conventional
droplet model to the specific case of a superconducting wire.
In this approach, one identifies a critical droplet as a saddle point solution to
the classical equations of motion (in our case, the GL equation). Dynamics along 
the negative mode
of this saddle point is supposed to mediate a transition between states that differ 
by one unit of the winding number.

On a closer inspection, however, one notices that the order parameter does not vanish
on the LAMH saddle point, except for initial states with exactly zero current. This
is because the  only current present in the GL theory is the London supercurrent
$I_s$, which is proportional to $|\Psi|^2$. For a solution with a {\em static} 
superfluid density, $I_s$ is the same everywhere along the wire, 
so, at a nonzero $I_s$, $\Psi$ cannot 
vanish at any point. Since vanishing of $\Psi$ is necessary for the phase to unwind, 
it is by no means obvious that the LAMH saddle point actually mediates transitions
between different winding numbers. One may equally well imagine that the system
starts at a winding number $W$, reaches the LAMH 
saddle point, and then rolls down along the negative mode, but remains all the time
in the sector with the same winding number.

Therefore, one is justified in looking for fluctuations (with {\em time-dependent} 
densities) that have a zero of $\Psi$ and thus allow the phase to 
unwind. Such fluctuations have been observed in numerical studies of superfluids 
flowing past obstacles \cite{obstacles}, see also Ref. \cite{AP}. 
The studies of Ref. \cite{obstacles} have used boundary conditions 
corresponding to a fixed value of superfluid velocity away from the obstacle. 
Here we want to consider transitions between states with different values of  
superfluid velocity and so use periodic boundary conditions (the ring geometry). 
In this case, the winding number in the vicinity of 
a uniform supercurrent state must be an integer. 

In a nearly uniform Bose fluid, a natural candidate for the critical fluctuation is 
one of the many time-dependent solutions (solitons \cite{solitons}) of the perfectly 
uniform problem. According to the preceding discussion,
we need a field for which the order parameter is completely extinguished at some
point. For a large enough length $L$ of the ring, a solution having this property
is
\be
\Psi(x, t) = \Psi_1 e^{-i\mu t + i m v_1 x} \tanh\frac{x - v_1 t}{2\xi_1} \; .
\label{moving}
\ee
Here $\Psi_1$ and $\mu$ are suitably chosen constants ($\mu$ is real, $\Psi_1$
is complex), $v_1 = (2\pi/mL) (W + \half)$, $W$ is the winding number of a nearby 
uniform state
(an integer), $m$ is the mass of the particle, and $\xi_1$ is the ``healing'' 
length, $\xi_1^{-2} = 4 gm |\Psi_1|^2$, $g$ is the coupling constant.

The field (\ref{moving}) solves the time-dependent Gross-Pitaevskii (GP)
equation at zero
external potential, and represents a depletion of density moving at velocity
$v_1$. 
The classical energy of this state is computed in the Appendix.
The momentum of this state is $m v_1 N$, 
where $N$ is the total number of particles. Relative 
to the uniform state with winding $W$ and superfluid velocity 
$v_0 = 2\pi W / mL$, the momentum is 
\be
\Delta P = mv_1N - mv_0 N = \pi n \; ,
\label{DP}
\ee
where $n = N/L$ is the linear density. In the limit 
$L\to \infty$ with $n$ and $W$ fixed, 
the field (\ref{moving}) becomes the classical counterpart 
of the extremal point of the Lieb-Liniger spectrum \cite{Lieb&Liniger} 
of the corresponding quantum problem---a Bose gas with a delta-function repulsion.

Note that, while in the limit of small $v_1$ the solution
(\ref{moving}) is in a certain sense
close to the corresponding LAMH saddle point, for larger $v_1$
the two solutions are substantially different.

We want to identify (\ref{moving}) as a critical
droplet mediating transitions between supercurrent states with winding numbers 
$W$ and $W+1$, i.e., superfluid velocities $v_0$ and
$v_2= 2\pi (W + 1)/ mL$. This identification shows immediately
that in a nearly uniform superfluid TAPS may be strongly suppressed.
Indeed, if the droplet is to decay into a state with winding $W$, the momentum
$\Delta P$ needs to be transferred to another excitation branch, i.e., phonons,
or to some external system.
A look at the Lieb-Liniger spectrum tells us that any phonon state with momentum
$\Delta P$ has energy larger than the energy of the droplet state.

The nearly uniform limit can be realized in a variety of physical systems. 
An interesting
and perhaps somewhat unexpected realization is a thin superconducting wire.
In this case, the analog of Bogoliubov's phonon is the gapless plasmon mode
\cite{Kulik&MS}, describing fluctuations of the superconducting density.
The large-scale (``hydrodynamic'') effective theory of these plasmons is a GP theory,
with the coupling constant $g=4e^2 / C$, where $C$ is the capacitance of the wire per
unit length (see, e.g., Ref. \cite{slip}). This description of superconductors
is specific
to one dimension, and the reason why it holds is that in 1D the screening of 
charges is weak,
so the large-scale dynamics is dominated by the charging energy, rather than 
the energy of condensation (as in the GL theory).
The order parameter $\Psi$ of this description is proportional (but not necessarily
equal) to 
the GL order parameter. It can be interpreted as the field of
Cooper pairs, and the entire description applies only at length scales larger than
the ``size'' of a pair, i.e., the GL coherence length $\xi_{GL}$. 
Since in practice $\xi_{GL}$ is much larger than the ``healing'' length $\xi$ 
obtained from the GP description, the size of a critical droplet will now be determined 
by $\xi_{GL}$, rather than $\xi$. Nevertheless, as we discuss
further in the concluding section, it is possible to adapt some of our results
to this case. In particular, we expect that at scales of order $\xi_{GL}$
disorder, typically present in superconducting wires,
self-averages, so that as an initial approximation we can use the uniform
limit with suitably renormalized parameters. At the next level of accuracy, we 
will need to include momentum transfer to disorder via various mechanisms.

Another possible realization of the nearly uniform limit is atomic 
superfluids---specifically,
trapped Bose gases---which can in principle be prepared is such a way
that there are no significant sources of disorder on scales shorter 
than the healing length.
In this case, the above considerations suggest that the requirement of
momentum transfer can present a major bottleneck for TAPS.

We want to stress that, as shown in Refs.
\cite{slip,P}, in {\em quantum} theory phase slips occur even in a perfectly 
uniform superfluid,
by thermally-assisted tunneling with momentum transfer to phonons.
The question we address in the present paper is whether phase slips are possible 
by a {\em classical} mechanism, such as thermal activation.
This question is particularly
relevant at temperatures $T \agt gn$ ($g$ is the coupling constant), where the
approximate methods used in Refs. \cite{slip,P} do not apply. 

Here,
we attempt to simulate TAPS numerically, using classical equations of motion.
Our simulations are microcanonical, i.e., there is no coupling to any external
heat bath, and are similar in spirit to simulations of sphaleron transitions
in the (1+1)-dimensional Abelian Higgs model \cite{GRS}. In our case, however,
because of the exact solvability of the theory at zero potential \cite{ZS}, 
we take 
special measures to ensure appropriate population of the phase space (see below).

Our main result is the complete
absence, within the classical mechanics, of TAPS in the uniform Bose gas. 
This leaves quantum tunneling as the only mechanism of supercurrent decay in 
this case.

We have found that TAPS appear in the presence
of an external potential $V(x)$, but at a rate that depends sensitively
on the magnitude and 
scale of variation of $V$. When $V$ is smooth and not too large, the rate is much 
smaller than that derived from the LAMH theory.
Fluctuations that mediate TAPS in such a smooth potential are
similar to the field (\ref{moving}). 

The paper is organized as follows. Details of the numerical procedure are described
in Sects. \ref{sect:init} (initial conditions) and \ref{sect:method} 
(the evolution algorithm). Numerical results are presented in Sect. \ref{sect:res}.
Our conclusions are summarized in Sect. \ref{sect:concl}, where we also discuss some
applications of our results.

\section{Initial conditions} \label{sect:init}
For the evolution equation, we have used the GP equation
\be
i \frac{\partial \Psi}{\partial t} = -\frac{1}{2m} \partial_x^2 \Psi +
g |\Psi|^2 \Psi + V(x) \Psi
\label{GP}
\ee
with periodic boundary conditions
\be
\Psi(x+L) = \Psi(x) \; .
\label{bc}
\ee
Here $\Psi(x,t)$ is the complex order parameter, $m$ is the mass of the particles,
$g > 0$ is the coupling constant, and $V(x)$ is an external potential. 
For the uniform case, $V(x)=0$. We have set $\hbar = 1$.

Important parameters that we will often use in what follows are
the zero-temperature speed of Bogoliubov's phonons
\be
c_0 = (g\nave / m)^{1/2} 
\label{c_0}
\ee
and the zero-temperature ``healing'' length
\be
\xi = (4 gm\nave)^{-1/2} \; ;
\label{xi}
\ee
$\nave$ is the average density of the gas. In a classical simulation, 
$\nave$ is obtained as the average of the density
$n = \Psi^\dagger \Psi$ over the entire lattice:
\be
\nave = \frac{1}{L} \int dx \Psi^\dagger \Psi \; .
\label{nave}
\ee

A precise characterization of temperature was not a goal of the present work.
This allowed us to initialize the field directly from the phonon power spectrum.
Namely, suppose we separate the initial field $\Psi$ into the modulus (square root 
of the density) and the phase:
\be
\Psi(x,t_i) = \sqrt{n(x)} e^{i\theta(x)} \; .
\label{separ}
\ee
For the phase, we write
\be
\theta(x) = \frac{2\pi W_i x}{L} + \theta_0 + \theta_1(x) \; , \
\label{wind}
\ee
where $W_i$ is the initial winding number, $\theta_0$ is an uninteresting constant,
which will be set to zero, and $\theta_1$ is a fluctuation. 
For the density, we write
\be
n(x) = \nave + \delta n(x) \; ,
\label{dndef}
\ee
where $\nave$ is the uniform average density, which we specify,
and $\delta n$ is a fluctuation. 

Next, the fluctuations $\theta_1$ and $\delta n$ 
are expressed using Bogoliubov's transformation
\ba
\theta_1(x)  & = & {1\over \sqrt{L}} 
\sum_{k \neq 0} \frac{\sqrt{Z_k}}{\sqrt{2\omega_k}} \left[
b_k  + b^\dagger_{-k} \right] e^{ik x} , \label{theta} \\
\delta n(x) & = & {i\over \sqrt{L}} 
\sum_{k \neq 0} \sqrt{\omega_k \over 2 Z_k} \left[
b_k - b^\dagger_{-k} \right] e^{ik x} , \label{dn}
\ea
where 
\ba
\omega_k & = & c_0 [k^2 + k^4 \xi^2]^{1/2} \; , \\
Z_k & = & g (1 + k^2 \xi^2) \; . \label{Z}
\ea
In our classical simulation,
we take $b_k$ and $b_k^\dagger$ to be random classical variables, whose moduli are
given by the Bose distribution
\be
|b_k|^2 = (e^{\Omega_k / T} - 1)^{-1} \; ,
\label{occup}
\ee
where 
\be
\Omega_k = \omega_k  + v k \; .
\label{Omega_k}
\ee
is the quasiparticle dispersion law in the presence of a superfluid velocity $v$.
The phase of $b_k$ is a (pseudo)random number uniformly
distributed between 0 and $2\pi$. 

In numerical work, instead of (\ref{separ}) we use the linearized version
\be
\Psi = {\cal N} \sqrt{\nave} \left( 1 + \frac{\delta n}{2\nave} \right) e^{i\theta} 
\; ,
\label{recon}
\ee
where ${\cal N}$ is a normalization coefficient enforcing the condition 
(\ref{nave}).

The use of the linearized Eq. (\ref{recon}) is consistent with the assumption
$\delta n \ll \nave$ of Bogoliubov's theory and with using 
the ideal-gas distribution
(\ref{occup}) for phonons. But how far is the result from the true equilibrium 
of the interacting system and how much should we worry about the difference? 

To begin with, notice that for a classical field the notion of a ``true'' 
equilibrium applies only to the low-frequency component, $\Omega_k \ll T$,
which is more or less classical. For a TAPS, we expect $k \sim 1/\xi$, i.e.,
$\Omega_k \sim g \nave$, so that at $T \gg g\nave$ a TAPS is essentially a 
classical fluctuation. For high-frequency modes, the classical approximation
is not parametrically justified, but under the above condition on the temperature
these only provide a ``heat bath'', and we do not expect the results to depend
very sensitively on the model we choose for them.

By the same token, it seems extremely unlikely that modest deviations 
from thermal
equilibrium can jeopardize the qualitative results of this work (such as
vanishing of the TAPS rate in the uniform system).

One may, however, raise the objection that, in the uniform case or for 
a weak potential, our way of generating the initial conditions
may not populate adequately the Lieb-Liniger branch, to which
for instance the state (\ref{moving}) belongs. To counter this objection, we
subject the field, after it has been initialized as described above, 
to a ``premixing'' stage,
wherein it evolves for a while in a relatively strong potential. After that,
the potential is switched to the desired strength, the winding number is
reset to the original value, and the actual evolution begins.
Premixing causes further deviations from thermal equilibrium but these are 
again deemed harmless for the reason stated above.

Two further comments are in order:

(i) Bogoliubov's theory  \cite{Bogoliubov}, when formulated in terms of density and
phase, as in (\ref{theta}) and (\ref{dn}), requires smallness of $\delta n$ but
does not require either smallness of $\theta_1$ or 
a non-vanishing expectation of the order parameter. It is therefore applicable 
even in 1D, cf. Ref. \cite{Popov}.

(ii) Within that theory, the properly subtracted variance of the density is
\be
\langle \delta n^2 \rangle_{T} - \langle \delta n^2 \rangle_{T=0} =
\frac{c_0}{2\pi g} \int_{-\infty}^{\infty} \frac{ dk |k|}{\sqrt{1 + k^2 \xi^2}}
\frac{1}{e^{\Omega_k/T} - 1} \; ,
\label{varn}
\ee
Of main interest to us here
is the temperature region $T \agt g\nave$. For an estimate of
$\delta n$, we will assume the limit
\be
T \gg \frac{c_0}{\xi} = 2 g\nave \; .
\label{high-T}
\ee
In this limit, the integral in (\ref{varn}) is saturated 
at small $k$: $k \sim \xi^{-1} (1-v^2/c_0^2)^{1/2}$ (this infrared sensitivity 
is specific to one dimension). At these $k$, 
the Bose distribution can be replaced by the classical
$T / \Omega_k$, and upon that replacement the integral can be easily evaluated:
\be
\langle \delta n^2 \rangle_{T} - \langle \delta n^2 \rangle_{T=0} \approx 
\frac{\nave T}{\sqrt{c_0^2 - v^2}} \; .
\label{varn2}
\ee
We conclude that (unless $v$ is very close to $c_0$) the smallness of a typical
density fluctuation, $\delta n / \nave \ll 1$, requires that
\be
T \ll c_0 \nave \; .
\label{Tcond}
\ee
This condition is compatible with (\ref{high-T}),
provided the dimensionless coupling $g / c_0$ is chosen small.

\section{Evolution algorithm} \label{sect:method}
In the code, it is convenient to work with rescaled, dimensionless variables, since this
reduces the number of parameters that need to be input. A natural unit of length is
the healing length $\xi$, and of time---the ratio
\be
\frac{\xi}{c_0} = \frac{1}{2g\nave} \; .
\label{t_unit}
\ee
So, we rescale variables in the GP equation (\ref{GP}) as follows:
\ba
x & \to & x / \xi \; , \label{rescx} \\
t & \to & 2g\nave t \; , \label{resct} \\
V & \to & V / 2g\nave \; . \label{rescV}
\ea
In this section, and in the plots, we use the same letter for a rescaled quantity as
in the rest of the paper for the original one. We also rescale the field $\Psi$ into
\be
\tpsi = \Psi / \sqrt{\nave} \; ,
\label{tpsi}
\ee
so that the average rescaled density is equal to 1: 
$(1/L) \int \tpsi^\dagger \tpsi dx = 1$.

In the rescaled variables, the GP equation becomes
\be
i \frac{\partial \tpsi}{\partial t} = - \partial_x^2 \tpsi +
\half |\tpsi|^2 \tpsi + V(x) \tpsi \; ,
\label{tGP}
\ee

For the purpose of numerical evolution, we associate to the right-hand side
of (\ref{tGP}) an operator $U$ that updates the field $\psi_n(x)$ at the 
$n$-th time step:
\be
\tpsi_{n+1}(x) = U[\tpsi_n(x), \dt] \; ;
\label{update}
\ee
$\dt = t_{n+1} - t_n$. For this work, we have used an operator-splitting algorithm,
in which the update is made in three steps:
\be
U[\tpsi(x), \dt] = U_1[U_2[U_1[\tpsi(x), \half \dt], \dt], \half \dt] \; .
\label{3steps}
\ee
where $U_1$ corresponds to the equation
\be
i\frac{\partial \tpsi}{\partial t} = \half |\tpsi|^2 \tpsi + V(x) \tpsi \; ,
\label{eq1}
\ee
and $U_2$ to the equation
\be
i\frac{\partial \tpsi}{\partial t} = - \partial_x^2 \tpsi \; .
\label{eq2}
\ee

To solve these equations, the field was discretized on a spatial lattice with
a uniform step $\dx$ and periodic boundary conditions. Eq. (\ref{eq1}) is local
in space, so $U_1$ can be defined site-by-site:
\be
U_1[\tpsi(x), \dt] = \prod_{j=0}^{M-1} u_j(\tpsi_j, \dt) \; .
\label{U1}
\ee
We have used
\be
u_j(\tpsi, \dt) = \frac{1-\sigma_j}{1+\sigma_j} \tpsi \; ,
\label{u_j}
\ee
with $\sigma_j = {i\over 2} [\half |\psi|^2 + V(x_j)] \dt$. The operator
$U_2$ is non-local in $x$-space but local in $k$-space; we have used 
\be
U_2[\tpsi(x), \dt] = \prod_{k=0}^{M-1} v_k(\tpsi^F_k, \dt) \; ,
\label{U2}
\ee
where $\tpsi^F$ is the fast Fourier transform (FFT) of $\tpsi$, and
\be
v_k(\tpsi^F, \dt) = \frac{1-\rho_k}{1+\rho_k} \tpsi^F
\label{v_k}
\ee
with $\rho_k = \half i k^2$. So, there are two FFTs (one direct and one inverse)
at each time step.

The resulting algorithm has second-order accuracy in time. It conserves the
number of particles exactly, while the energy non-conservation is controlled
by the time-step $\dt$.

\section{Numerical results} \label{sect:res}
We present results from simulations on an $M=1024$ lattice with 
the following values of the dimensionless parameters:
$L/\xi =79$, $T/2g\nave=1.2$, and $g/c_0 = 0.1$ (a weak coupling).
We have used the periodic potential
\be
V(x) = V_0 \sin (2\pi x / l) \; .
\label{V}
\ee
For periodicity, me must have $l = L/q$, where $q$ is an integer. We have used
$q=15$, so that $l/\xi \approx 5.3$. ``Premixing'' was done with 
$V_0/2g\nave=0.3$ and
lasted for $\Delta t = 200$ (in our dimensionless units). The premixing stage 
is not shown in the plots, i.e., $t=0$ corresponds to the beginning of 
the actual evolution with a smaller potential. 
The winding number changes during premixing,
but at $t=0$ it is reset to the original value $W_i=3$.

\begin{figure}
\leavevmode\epsfxsize=3.25in \epsfbox{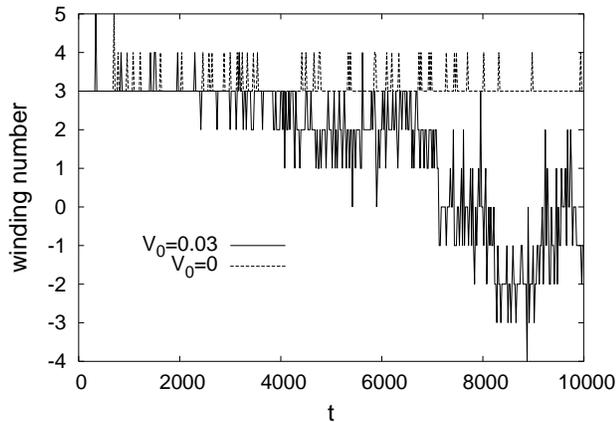}
\vspace*{0.2in}
\caption{Evolution of the winding number for $V_0/2g\nave=0.03$ and $V_0=0.$}
\label{fig:wind}
\end{figure}
In Fig. \ref{fig:wind}, we show the evolution of the winding number with 
and without a potential. 
In the uniform case ($V_0=0$), the winding number fluctuates but always
quickly returns
back to the initial value $W=3$. There is no overall relaxation. This behavior 
persisted as we went to larger values of the temperature. Based on these and 
similar other results, we conclude that, in the uniform 1D superfluid, supercurrent 
is classically stable for any value of superfluid velocity satisfying Landau's
criterion. Thus, the only mechanism of supercurrent decay in this case is 
a quantum effect---the phonon-assisted tunneling considered in Refs. \cite{slip,P}.

In the presence of a (weak) potential, while there are still
many unsuccessful phase-slip attempts, the winding number eventually relaxes towards 
zero.

These results may be compared to those of numerical simulations of sphaleron
transition in the (1+1)-dimensional Abelian Higgs model \cite{GRS}. Sphalerons are
very similar to TAPS except that they connect states with zero current, so there 
is no issue of momentum transfer. In the simulations of Ref. \cite{GRS}, they
readily occur in the absence of any external potential.

\begin{figure}
\leavevmode\epsfxsize=3.25in \epsfbox{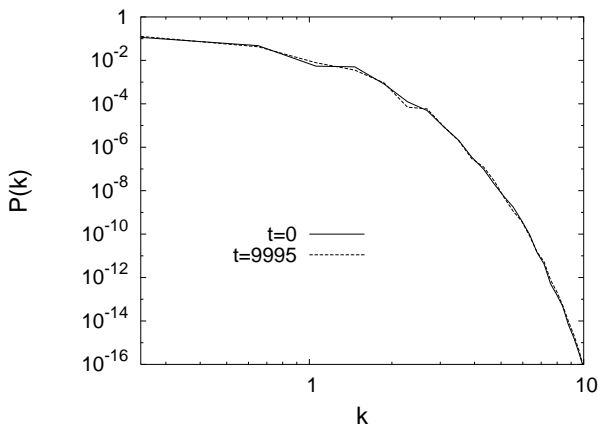}
\vspace*{0.2in}
\caption{Power spectrum of the field $\psi$ for $k>0$ in the uniform system
at the beginning and near the end of the simulation.}
\label{fig:pws}
\end{figure}
In Fig. \ref{fig:pws}, we show the power spectrum of $\psi$ in the case $V_0=0$ 
at the beginning and near the end of the simulation (only positive
$k$ are shown; results for $k \leq 0$ are similar). We see that the spectrum
exhibits remarkable stability. Although one expects that eventually fluctuations
will propagate to the ultraviolet (a manifestation of the Rayleigh-Jeans problem
of classical statistics), this clearly does not happen on the timescale of our
simulation.

\begin{figure}
\leavevmode\epsfxsize=3.25in \epsfbox{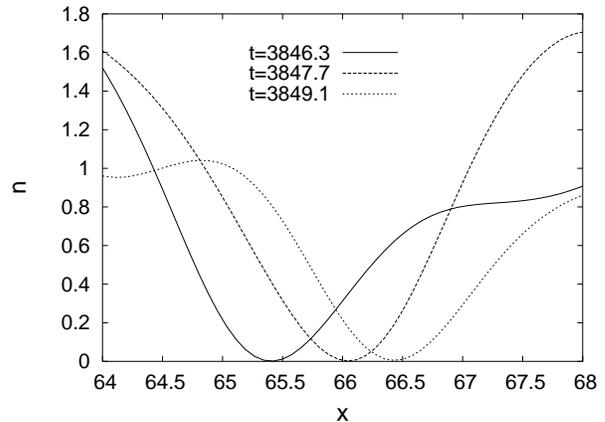}
\vspace*{0.2in}
\caption{The ``flying droplet'': a moving depletion of density that mediates 
a phase slip in the presence of a potential.}
\label{fig:fly}
\end{figure} 
In Fig. \ref{fig:fly}, we plot the density profiles of the fluctuation
mediating one of the phase slips in Fig. \ref{fig:wind} for $V_0 \neq 0$,
at three different moments of time. Comparing these profiles with 
Eq. (\ref{moving}),
we find that the spatial size of the droplet is about 2.5 times smaller than
predicted by that equation. This is not surprising given that, in the present
simulation, fluctuations of density are relatively large:
$\dn$ is about half of $\nave$. Since the typical wavelength of
these fluctuations is of the same order as the size of the droplet
[cf. comment (ii) in Sect. \ref{sect:init}], at $\delta n \sim \nave$ we loose the
notion of an effective 
long-range theory to which (\ref{moving}) could be a solution. 

This interpretation
is confirmed by going to smaller $T$ (and stronger potentials), for which
$\delta n /\nave$ is smaller, and the droplet 
size is expected to be closer to the value
predicted by Eq. (\ref{moving}). For example, for $T/2g\nave=0.9$ and 
$V_0/2g\nave=0.1$, the
difference reduces to a factor of about 1.6.

\section{Discussion}
\label{sect:concl}
The main result of this paper is that in a {\em uniform} 1D Bose gas 
{\em classical} phase slips are completely blocked out. We find this result
nontrivial and even surprising, given that momentum conservation, while
characteristic of the uniform system, by itself does not prohibit phase slips: 
momentum released from the supercurrent can be absorbed by phonons. Indeed, 
{\em quantum-mechanically}, phase slips are possible even in the perfectly
uniform system \cite{slip,P}. What we have shown here, then, is that quantum 
effects remain the only source of supercurrent decay in the uniform case.

Our results, both for the uniform system and in the presence of a potential,
are consistent with identifying the moving droplet (\ref{moving}) as the
fluctuation mediating thermal phase slips. As we have already noted, at nonzero
$v_1$ it is distinct from the LAMH saddle point. 

The nearly uniform limit can presumably be realized in trapped Bose gases.
Another system where, as already mentioned in the Introduction, the droplet 
(\ref{moving}) may play a role is a thin superconducting wire. In this case, 
the field $\Psi$ describes a fluid of Cooper pairs, and
for the reasons already indicated---the weak screening in 1D and the resulting
special role of the charging energy---the GP equation is well justified
for description of long-wavelength plasma oscillations---the gapless plasmon
mode \cite{Kulik&MS}.
However, this description breaks down at scales of order of the ``size'' of
a pair, i.e., the Ginzburg-Landau coherence length $\xi_{GL}$. Since in practice
$\xi_{GL}$ is much larger than the healing length $\xi$ obtained from
the GP description, the size of the critical droplet will now be determined
by $\xi_{GL}$, rather than $\xi$. In the following discussion, we will
need only the energy of droplet, which can be calculated using the Ginzburg-Landau
free energy. The latter differs from the GP Hamiltonian (\ref{H}) only by the
value of the quartic coupling $g$ and by the presence of a $|\Psi|^2$ term,
which in the GP case would correspond to a chemical potential. Although the
activation energies 
(\ref{diff1}), (\ref{diff2}) are computed in the Appendix at a fixed number of 
particles, the same expressions also apply at a fixed chemical potential.
As for the coupling $g$, it will be sufficient to consider it 
as a phenomenological parameter. This allows us to apply the expressions
obtained in the Appendix to the case of a superconducting wire. 

Since a moving droplet can now transfer momentum to 
normal electrons and, through them, to the disorder potential, we expect that
for sufficiently strong disorder the TAPS rate
is determined simply by the droplet's activation energy. The activation energies
for transitions decreasing and increasing the winding number by one unit
are given by Eqs. (\ref{diff1}), (\ref{diff2}). Unfortunately,
the form of the $I$--$V$ curve obtained from these expressions differs from that 
obtained in the LAMH theory only for currents comparable to the GL critical 
current. This makes it difficult to draw an experimental distinction between the
activation energies (\ref{diff1}), (\ref{diff2}) and their LAMH counterparts.
If, at smaller currents, we regard the last term in 
Eqs. (\ref{diff1}), (\ref{diff2}) as current-independent, the voltage drop found
from these equations is proportional to $\sinh (I_s/I_0)$ 
where $I_s$ is the supercurrent, and $I_0= 2e T /\pi$. This form of the nonlinear
$I$--$V$ curve, the same as in the LAMH theory in the equivalent limit \cite{LAMH}, 
has been
recently found to be in good agreement with experiment \cite{Rogachev&al}.

Finally, we mention one technical result of our work, namely,
the extent to which it turned out possible to simulate thermal field theory 
classically. We refer here to the remarkable stability of the classical
power spectrum against
spreading towards the ultraviolet, cf. Fig. \ref{fig:pws}. This is similar
to behavior observed in three dimensions, in simulations of highly {\em nonthermal} 
states produced by parametric amplification \cite{Khlebnikov:1996mc}.
One may wonder if this property holds more generally, so that the classical model 
of a thermal state we have used here 
can be applied to other (weakly-coupled) systems
where the high-frequency modes provide an internal ``heat bath'' for 
the low-frequency, classical component.

After this work was completed, we have learned of a paper by Polkovnikov 
{\em et al.} \cite{Polkovnikov&al} who studied numerically classical decay of
supercurrent in an optical lattice. Their calculation corresponds to a strong 
periodic potential---the limit opposite to ours.

The author thanks V. Ambegaokar, L. Pitaevskii, A. Rogachev, and 
A. Zubarev for useful discussions.
This work was supported 
in part by the U.S. Department of Energy through Grant DE-FG02-91ER40681 
(Task B).

\appendix
\section{Energy of a moving droplet}
Here we compute the energy of the field (\ref{moving}), relative to the
energy of the neighboring uniform states: one with winding number $W$ and superfluid
velocity $v_0 = 2\pi W/ mL$, and the other with winding $W+1$ and superfluid velocity
$v_2 = 2\pi (W + 1)/ mL$.

We start with the Hamiltonian corresponding to Eq. (\ref{GP}) with $V = 0$,
\be
H_0 =   \int dx \left( \frac{1}{2m} |\partial_x \Psi|^2 + \frac{g}{2} |\Psi|^4 
\right) \; ,
\label{H}
\ee
and compute the energy at a fixed total number of particles $N$. So, 
there is no need to include a chemical potential.

Substituting (\ref{moving}) into (\ref{H}) and integrating over $x$, we obtain
(to exponential accuracy in $\xi_1/L$)
\be
E_1 = \frac{m}{2} v_1^2 N  + \frac{g}{2} |\Psi_1|^4 (L - \frac{8}{3} \xi_1) \; ,
\label{E_1}
\ee
where $\xi_1$ is the healing length defined after Eq. (\ref{moving}).  
The particle number is
\be
N = \int dx \Psi^\dagger \Psi = (L - 4\xi_1) |\Psi_1|^2 \; .
\label{N}
\ee
The energies of the neighboring uniform states are
\be
E_{0,2} = \frac{m}{2} v_{0,2}^2 N + \frac{g}{2} |\Psi_0|^4 L \; ,
\label{E_0}
\ee
where $|\Psi_0|^2 = N/L$. To the leading order in $1/L$, $\xi_1 \approx \xi$,
\ba
v_1^2 - v_2^2 & \approx &  -\frac{2\pi v_1}{mL}  \; , \\
|\Psi_1|^2 - |\Psi_0|^2 & \approx & \frac{4\xi}{L} |\Psi_0|^2 
\ea
(we allow for the possibility that $W \propto L$). Then,
\ba
E_1 - E_0 \approx \pi v_1 |\Psi_0|^2 + \frac{8}{3} g \xi |\Psi_0|^4 \; , 
\label{diff1} \\
E_1 - E_2 \approx - \pi v_1 |\Psi_0|^2 + \frac{8}{3} g \xi |\Psi_0|^4 \; .
\label{diff2}
\ea
These expressions coincide with the LAMH activation energies \cite{LAMH} expanded
to the first order in supercurrent.

\end{document}